\documentclass{pasj00}
\draft

\begin{document}
\SetRunningHead{F. Usui et al.}{Infrared Asteroid Surveys: IRAS, AKARI, and WISE}

\title{A Comparative Study of Infrared Asteroid Surveys: IRAS, AKARI, and WISE}

\author{
Fumihiko  Usui\altaffilmark{1}, 
Sunao     Hasegawa\altaffilmark{2}, 
Masateru  Ishiguro\altaffilmark{3}, 
Thomas G. M\"{u}ller\altaffilmark{4}, 
and
Takafumi  Ootsubo\altaffilmark{5},  
}

\altaffiltext{1}{Department of Astronomy, Graduate School of Science, 
The University of Tokyo, 7-3-1 Hongo, Bunkyo-ku, Tokyo 113-0033}
\email{usui@astron.s.u-tokyo.ac.jp}

\altaffiltext{2}{Institute of Space and Astronautical Science, 
Japan Aerospace Exploration Agency, 
3-1-1 Yoshinodai, Chuo-ku, Sagamihara 252-5210}

\altaffiltext{3}{Department of Physics and Astronomy, 
Seoul National University, 
San 56-1, Shillim-dong Gwanak-gu, Seoul 151-742, South Korea}

\altaffiltext{4}{Max-Planck-Institut f\"{u}r Extraterrestrische Physik, 
Giessenbachstra\ss e, 85748 Garching, Germany}

\altaffiltext{5}{Astronomical Institute, Tohoku University, 
6-3 Aoba, Aramaki, Aoba-ku, Sendai 980-8578}


\maketitle

\begin{abstract}
We present a comparative study of three infrared asteroid surveys 
based on the size and albedo data from the 
Infrared Astronomical Satellite (IRAS), 
the Japanese infrared satellite AKARI, and 
the Wide-field Infrared Survey Explorer (WISE). 
Our study showed that: 
(i) the total number of asteroids detected with diameter and albedo information with these three surveyors is 138,285, 
which is largely contributed by WISE; 
(ii) the diameters and albedos measured by the three surveyors for 1,993 commonly detected asteroids 
are in good agreement, 
and within $\pm$10\% in diameter and $\pm$22\% in albedo at 1$\sigma$ deviation level.
It is true that WISE offers size and albedo of a large fraction ($>20$\%) of known asteroids 
down to a few km bodies, but we would suggest that the IRAS and AKARI catalogs compensate 
for larger asteroids up to several hundred km, especially in the main belt region. 
We discuss the complementarity of these three catalogs in order to facilitate the use of these data sets 
for characterizing the physical properties of minor planets. 

\end{abstract}

\section{Introduction}

Presently, the number of asteroids is known to be more than 620,000. 
Most asteroids are, however, known only from their orbital data and their other properties are poorly 
constrained. In particular, size of asteroid, which is one of the most basic physical quantities, has been 
unknown for most asteroids.
Several techniques have been developed to determine
the size of asteroid. 
One of the most effective methods for measuring asteroidal size and albedo indirectly is through the use of radiometry, 
where a combination of the thermal infrared flux and the absolute magnitude as the reflected sunlight. 
This radiometric method can provide unique data for asteroidal size and albedo. 
Observations in mid-infrared wavelengths are suitable for studying asteroids with this method, particularly 
in the inner solar system inside the orbit of Jupiter. 
Using radiometric measurements, a large number of objects can be observed in a short period 
of time, providing coherent data for large populations of asteroids within the asteroid belt. 

Infrared observations can be made still better under ideal circumstances,  
from space. 
The first space-borne infrared telescope is 
the Infrared Astronomical Satellite (IRAS; ~\cite{Neugebauer1984}), launched in 1983 
and performed a survey of the entire sky. 
To date, there are two other infrared astronomical satellites dedicated to all-sky surveys: 
the Japanese infrared satellite AKARI~\citep{Murakami2007}, and 
the Wide-field Infrared Survey Explorer (WISE; ~\cite{Wright2010}). 
Other space-borne infrared telescopes,  
e.g., the Midcourse Space Experiment (MSX; \cite{Mill94}), 
the Infrared Space Observatory (ISO; \cite{Kessler96}), 
the Spitzer Space Telescope~\citep{Werner2004}, 
and the Herschel Space Observatory~\citep{Pilbratt2010}  
have conducted a series of observations with imaging and/or spectroscopy of asteroids. 
Based on the all-sky survey data obtained by IRAS, AKARI, and WISE (hereafter I--A--W), 
the largest asteroid catalogs containing the size and albedo data were 
constructed.
However, at present little is known about the consistency of these three catalogs of asteroidal data. 
Performance of the on-board detectors and the survey strategies are different, the time and season of the observations 
and the duration of surveys are different, and the thermal model of asteroids adopted for determining size and albedo 
are different, between I--A--W. 
The relationship between these three catalogs should be checked in order to facilitate the use of these data sets 
for scientific purposes.

In this paper, we compare the asteroidal catalog data obtained by I--A--W
to investigate the consistency and characteristics of these data sets and reveal some benefits of 
the usage of synthesized these three data for studying the physical properties of minor planets. 
We have reviewed each surveyor and its data set, and have compiled these data into 
a single data set. Subsequently, we compare the number and distribution of the asteroids 
detected by these satellites, 
and discuss the completeness of the data sets obtained from each of the three satellites.

\section{Size and albedo data set}

\subsection{Infrared all-sky surveyors: IRAS, AKARI, and WISE} 
In this paper the results obtained by IRAS, AKARI, and WISE (I--A--W) are 
used. Technical specifications of these three satellites are summarized in table~\ref{3 satellites}. 

A pioneering systematic asteroid survey with a space-borne 
telescope was made by IRAS in 1983.
IRAS performed a survey of the entire sky with a 57~cm aperture telescope in ten~months. 
The IRAS asteroid catalog was provided 
by \citet{Tedesco2002}
as the supplemental IRAS minor planet survey\footnote{http://sbn.psi.edu/pds/resource/imps.html} (SIMPS). 
The SIMPS includes the averaged results for 2,470 asteroids, 
which is 2,228 asteroids with at least two accepted observations 
and 242 that only have a single accepted sighting in a single band, 
by using the Standard Thermal Model (STM;~\cite{Lebofsky1986}). 
These data were revised by \citet{Ryan2010}, which used 
the Near-Earth Asteroid Thermal Model (NEATM;~\cite{Harris1998}) to derive asteroidal sizes and albedos, 
and which applied stricter criteria, 
selecting only objects with reported fluxes in three or four band-passes, 
and compiled the sizes and albedos for 1,483 objects. 

AKARI is a second-generation infrared all-sky surveyor following IRAS. 
AKARI conducted a 16-month survey 
in six wavelength bands from the mid- to far-infrared with a 68.5cm aperture telescope. 
From many images taken in the mid-infrared part of the All-Sky Survey with 
the Infrared Camera (IRC;~\cite{Onaka2007}) on board AKARI, 
the infrared signals from asteroids were extracted and 
the Asteroid Catalog Using AKARI\footnote{http://darts.jaxa.jp/ir/akari/catalogue/AcuA.html} (AcuA) was constructed, which 
contains the size and albedo data for 
5,120 asteroids~\citep{Usui2011}. 
Additionally, from the slow-scan observations in the pointed observation mode of AKARI, 
a serendipitous asteroidal catalog was constructed (AcuA-ISS; \cite{Hasegawa2013}), 
which includes 
data from 88 main belt asteroids. 

The WISE satellite, launched in December of 2009, made a mapping of the whole sky 
in four bands in the near- to mid- infrared, 
with a 40~cm 
aperture telescope. 
While IRAS and AKARI conducted continuous scanning of the sky with 
horizontally aligned detector arrays 
along the attitude control of the satellites 
at a constant scan rate (IRAS: \timeform{3'.855}~s$^{-1}$, AKARI: \timeform{3'.6}~s$^{-1}$), 
WISE used a different approach. 
The payload of WISE included a cryogenic scan mirror 
driven in a sawtooth pattern to cancel the orbital motion of the satellite, 
and to freeze the line of sight during 
each 11~s exposure interval 
(actual exposure times were 7.7~s in 3.4 and 4.6~\micron~bands and 8.8~s in 12 and 22~\micron~bands; 
\cite{Schwalm2005, Cutri2013}), 
which
achieves its wide field-of-view and high sensitivity. 
WISE accomplished its four-band full cryogenic survey phase in seven~months, and 
continued survey observations for about six months after its cryogenic tanks became empty. 
The WISE asteroid data set is the most recent and the largest asteroid catalog 
provided by an enhancement program called NEOWISE~\citep{Mainzer2011a}. 
The sizes and albedos of asteroids measured with WISE are currently published as a series of papers: 
\citet{Masiero2011, Masiero2012} for the main belt asteroids, 
\citet{Mainzer2011d,Mainzer2012} for the near-Earth asteroids, 
\citet{Grav2011, Grav2012a} for the Jovian Trojans, 
\citet{Grav2012b} for the Hilda group, 
and \citet{Bauer2013} for the scattered disk objects and Centaur populations. 

In total, there are 137,837 asteroids in the WISE data set 
with valid diameter and albedo information. 

\subsection{Comparison of IRAS, AKARI, and WISE data sets}
\label{Comparison of ternary data sets}

\begin{figure*}
{\FigureFile(150mm,80mm){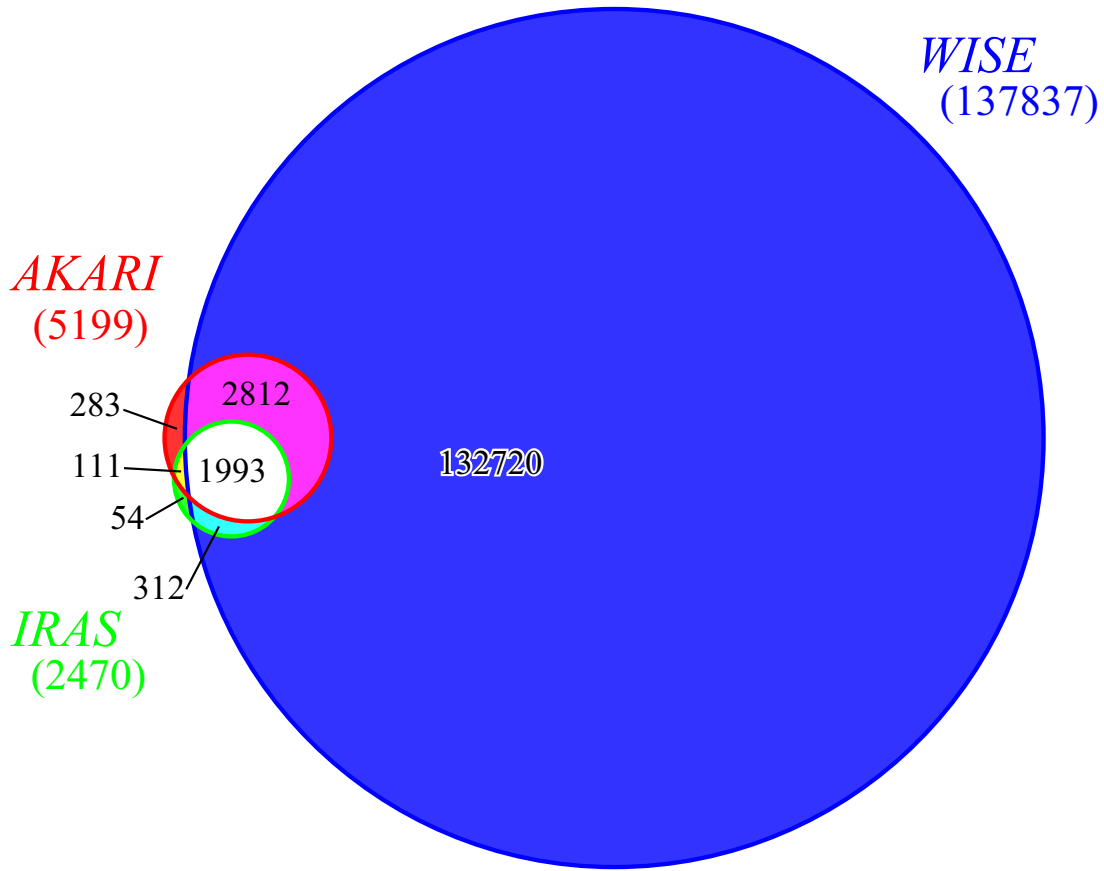}}
\caption{
A comparison of the number of asteroids detected with
{IRAS}, {AKARI}, and {WISE} shown as a Venn diagram.
The total number of asteroids detected with either {IRAS}, {AKARI}, or {WISE} is 138,285.
}
\label{Venn diagram}
\end{figure*}

Figure~\ref{Venn diagram} shows the relationship between the number of asteroids detected with each of these surveyors in 
the form of a Venn diagram. 
The number of asteroids detected with any three satellites is 138,285, which is 22\% of 
currently known asteroids with orbits, 
and all three satellites detected a common 1,993 asteroids. 
Most objects were only detected with {WISE}, 
because {WISE} is about two orders of magnitude more sensitive 
than {IRAS} or {AKARI} at mid-infrared wavelengths 
(table~\ref{3 satellites}). 
Nevertheless, 448 asteroids were detected with only {AKARI} and/or {IRAS}.

\begin{figure*}
{\FigureFile(150mm,80mm){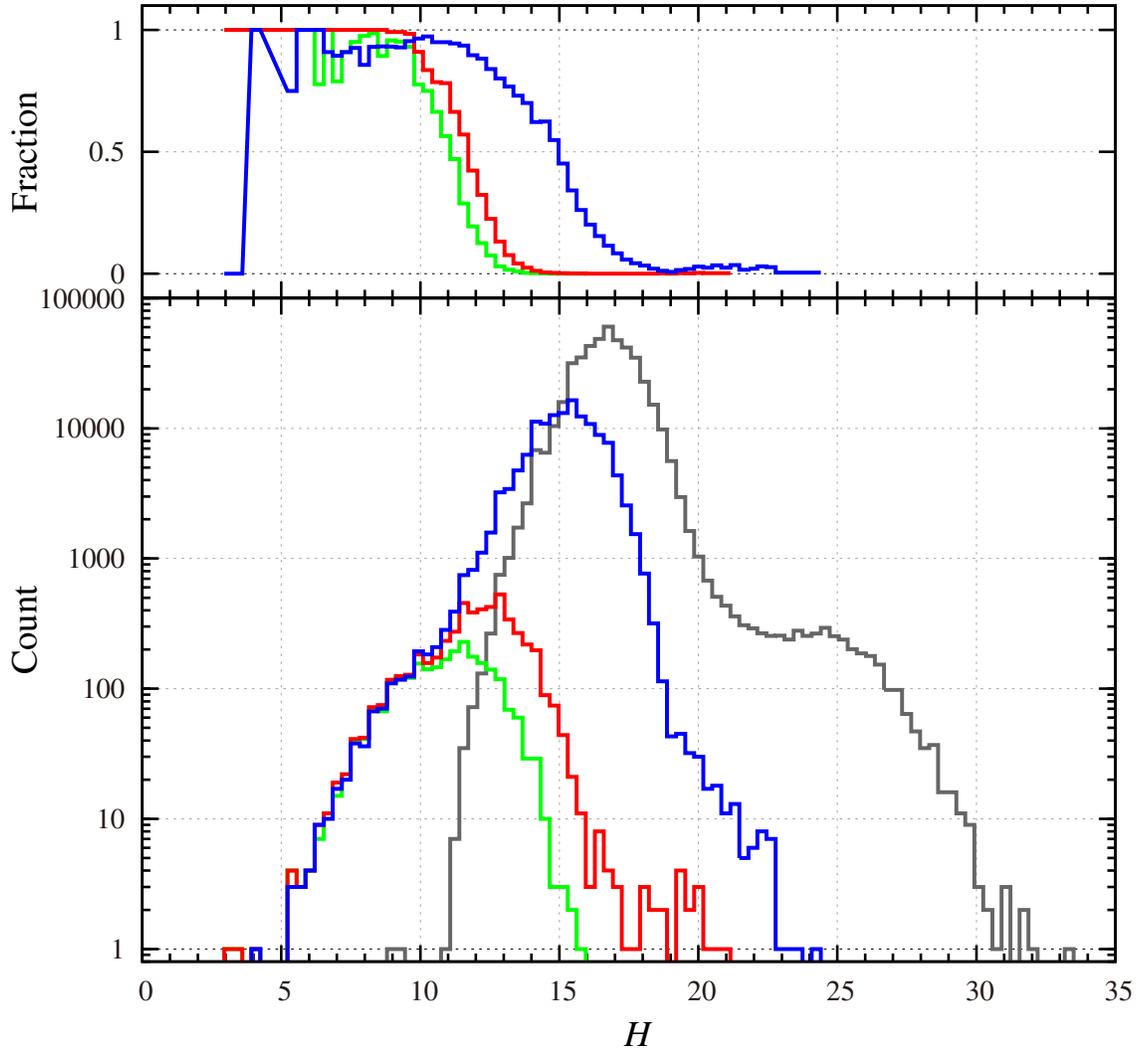}}
\caption{
Distribution of absolute magnitude ($H$) for asteroids detected with I--A--W, or 
undetected, with 
known orbits with semi-major axes smaller than 6 AU. 
The upper panel shows the fraction of detected asteroids of the total, and 
the lower panel shows the distribution of detected asteroids, 
by {IRAS} (green), {AKARI} (red), {WISE} (blue), 
and asteroids undetected by I--A--W (gray). 
Note that the absolute magnitude was not measured with these infrared surveyors, 
but determined by ground-based observational data in visible wavelengths 
(see, e.g., \cite{Bowell1994}). 
}
\label{Hmag distribution}
\end{figure*}

Figure~\ref{Hmag distribution} shows the distribution of the absolute magnitude for all asteroids detected 
by I--A--W with known orbits and semimajor axes smaller than 6 AU. 
This figure can be interpreted as reflecting 
the completeness of detections of known asteroids with the size and albedo data by three satellites. 
The {AKARI} asteroid catalog, which was constructed based on 16 months of the All-Sky Survey data,
provides a 100\% complete data set of all asteroids brighter than absolute magnitude of $H < 9$, 
which includes 40~km size asteroids at minimum
(and all main belt asteroids brighter than $H < 10.3$, corresponding to 20~km or larger objects; \cite{Usui2013}). 
It is important to include the entire population of asteroids with $H < 9$ for investigating 
the mass distribution of asteroids: 
asteroids with $H < 9$ account for more than 90\% of the total mass of all asteroids.
While {WISE} detected much smaller asteroids that peaked at $H~\sim~15$, 
with corresponding diameter $d~\geq~1.5$~km, 
some larger (and/or brighter) objects were not detected. 
It is notable that 
the numbers of undetected asteroids that were brighter than $H < 9$ are 
23 for {IRAS}, 1 for {AKARI}, and 33 for {WISE}, which 
are very small proportion of the total number of objects detected by each satellite. 
There is only one asteroids with $H < 9$ that do not have measurements of its size and albedo so far: 
1927~LA ($H$ = 8.81). 
This belongs to the outer main belt asteroids and has an expected size of $d = 77$~km, assuming albedo of $p_{\rm v} = 0.09$. 
The discovery of 1927~LA was reported in 1927 by Albrecht Kahrstedt at the Heidelberg-K\"{o}nigstuhl Observatory, Germany, 
but it was a single-apparition with only three observations, and one of them was noted as being in question 
(refer to Astronomische Nachrichten 232, 257 (1928) and also to the Minor Planet Center). 
No further follow-up observation has identified 1927~LA since these studies and, as such, we consider 
its existence is doubtful at present. 

\begin{figure*}
\begin{center}
\hspace*{-1cm}
\begin{tabular}{c}
{\FigureFile(100mm,80mm){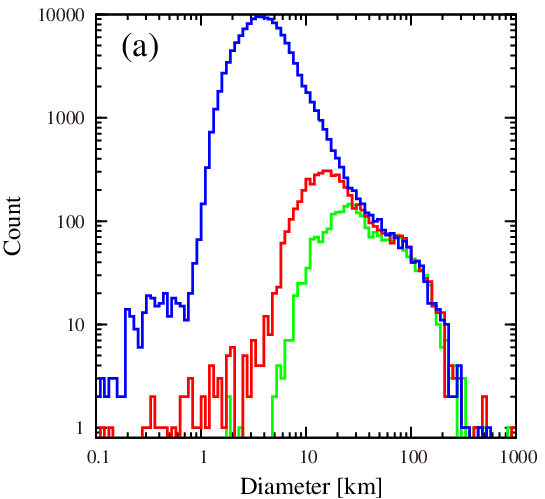}}\\
{\FigureFile(100mm,80mm){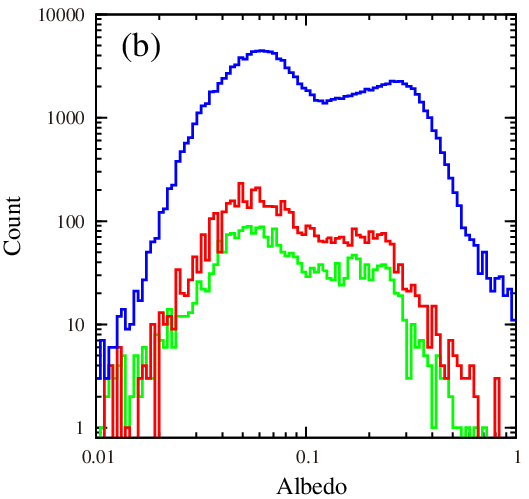}}\\
\end{tabular}
\end{center}
\caption{
Distribution of (a) diameter and (b) albedo of asteroids detected by 
IRAS (green), AKARI (red), and WISE (blue). 
The bin size is set at 100 segments for the range of 0.1 to 1000~km
in the logarithmic scale for (a) and 100 segments for the range of 0.01 to 1.0 
in the logarithmic scale for (b). 
}
\label{fig:histogram of each 3 catalog}
\end{figure*}

Figure~\ref{fig:histogram of each 3 catalog} illustrates the distribution of asteroids 
identified by I--A--W as a function of size and albedo. 
The size distribution in 
figure~\ref{fig:histogram of each 3 catalog}~(a) 
shows diameter maxima at ca. $d=25$~km, 14~km, and 3~km 
for IRAS, AKARI, and WISE, respectively. This suggests that 
asteroids larger than these limits are almost always detected by each satellite, 
but that the completeness of the catalog for each satellite decreases rapidly for asteroids smaller than these values. 
The albedo distribution in 
figure~\ref{fig:histogram of each 3 catalog}~(b) 
exhibits a bimodal distribution: 
the primary peak is found around the geometric albedo of $p_{\rm v} \sim 0.06$ for each data set; 
the secondary peaks are found at 0.16 for IRAS and AKARI, but at 0.25 for WISE. 
In these three data sets, there are extremely bright asteroids ($p_{\rm v} > 0.5$), which are 
all smaller than 55~km, and mostly smaller than 10~km, in the main belt as well as the near-Earth region. 
The numbers of these anomalously  high-albedo objects are, 5 in the IRAS catalog (0.20\% of its total), 
24 in the AKARI (0.46\%), and 888 in the WISE (0.64\%). 
It should be noticed that brighter high-albedo objects are more likely to be 
discovered, identified, and photometrically measured in visible wavelengths, especially 
toward smaller sized objects; in other words, the significant observational biases and 
selection effects still exist against darker and smaller objects. 
In contrast, infrared surveys are less biased against low albedo objects (e.g., \cite{Mainzer2011a}). 

\begin{figure*}
\begin{center}
\hspace*{-1cm}
\begin{tabular}{ccc}
\hspace{10mm}{\small (a) AKARI vs. IRAS} &%
\hspace{10mm}{\small (b) AKARI vs. WISE} &%
\hspace{10mm}{\small (c) WISE  vs. IRAS} \\
{\FigureFile(49mm,80mm){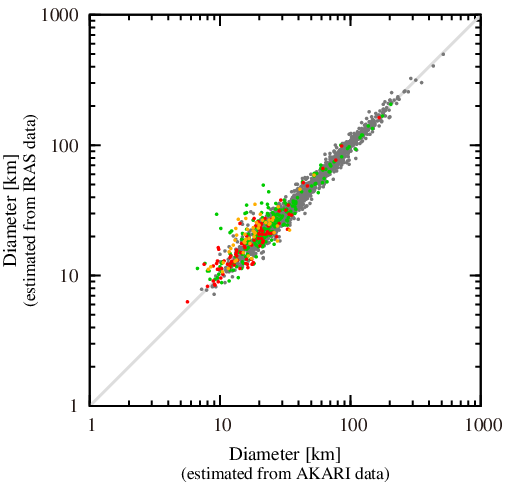}}&%
{\FigureFile(49mm,80mm){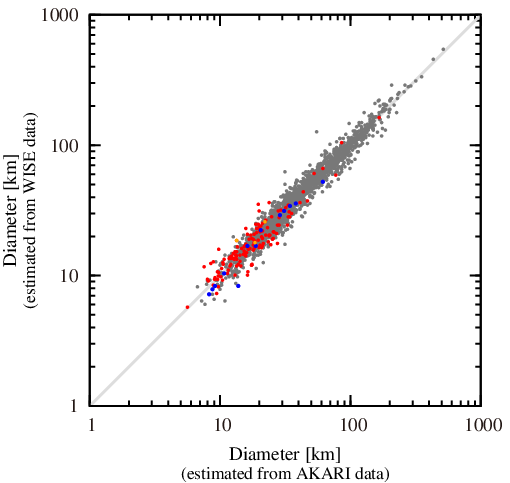}}&%
{\FigureFile(49mm,80mm){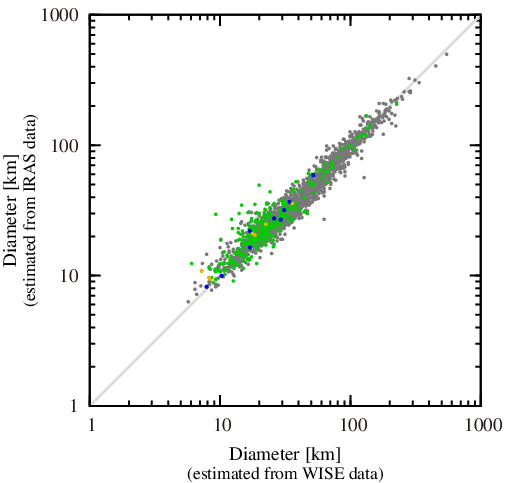}}\\
\\
\hspace{10mm}{\small (d) AKARI vs. IRAS} &%
\hspace{10mm}{\small (e) AKARI vs. WISE} &%
\hspace{10mm}{\small (f) WISE  vs. IRAS} \\
{\FigureFile(49mm,80mm){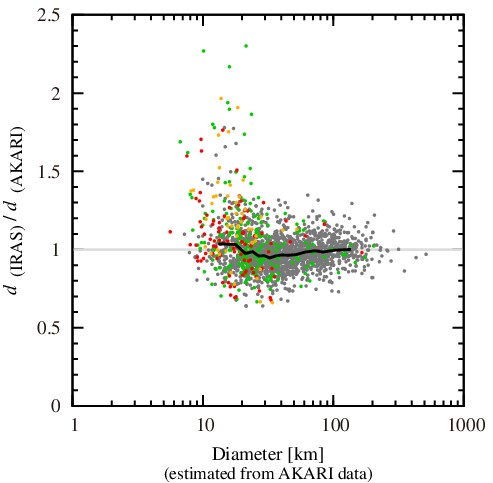}}&%
{\FigureFile(49mm,80mm){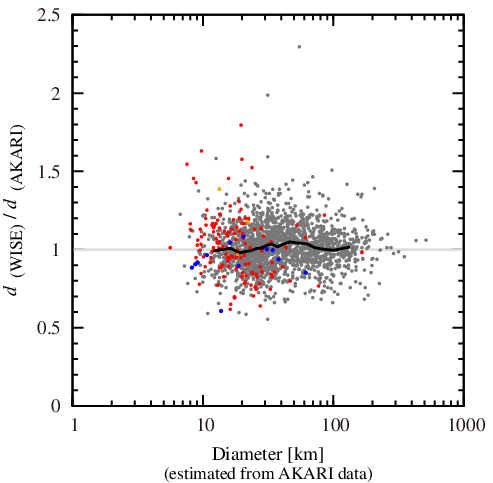}}&%
{\FigureFile(49mm,80mm){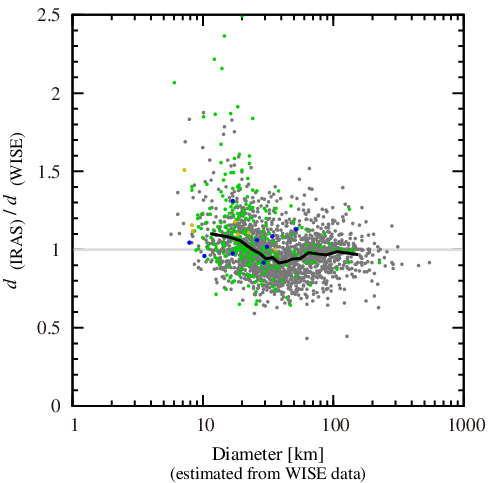}}\\
\end{tabular}
\end{center}
\caption{
A comparison of the differences in diameter obtained on the asteroids commonly detected by all three satellites, i.e.,
IRAS, AKARI, and WISE (1,993 asteroids). 
Green, red, and blue dots denote the asteroids observed with 
the single accepted sightings of IRAS~\citep{Tedesco2002}, 
the single event detections of AKARI~\citep{Usui2011}, 
and 
the single band detections of WISE~\citep{Masiero2011}, respectively. 
Yellow dots denote the asteroids with the single observations in both satellites in each panel. 
Note that in the WISE data any objects with a single detection in a single band are not accepted
(see, e.g., \cite{Mainzer2011a}); 
the data with at least four time detections in a single band are included. 
Apart from these small number of detections, 
the mean values of the diameter ratios obtained by the different satellites are 
0.982 (0.113), 
1.013 (0.139), 
and 
0.982 (0.154) 
for (d), (e), and (f), respectively, where the numbers in parentheses 
are the standard deviation (1$\sigma$). 
Thick black lines in the lower three panels show the running averages of the diameters 
for 100 object-wide bins, excluding the data corresponding to the small number of detections. 
}
\label{fig: comparing diameter}
\end{figure*}

\begin{figure*}
\begin{center}
\hspace*{-1cm}
\begin{tabular}{ccc}
\hspace{10mm}{\small (a) AKARI vs. IRAS} &%
\hspace{10mm}{\small (b) AKARI vs. WISE} &%
\hspace{10mm}{\small (c) WISE  vs. IRAS} \\
{\FigureFile(49mm,80mm){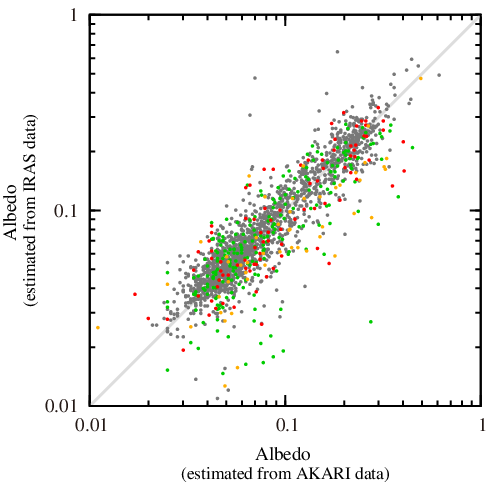}}&%
{\FigureFile(49mm,80mm){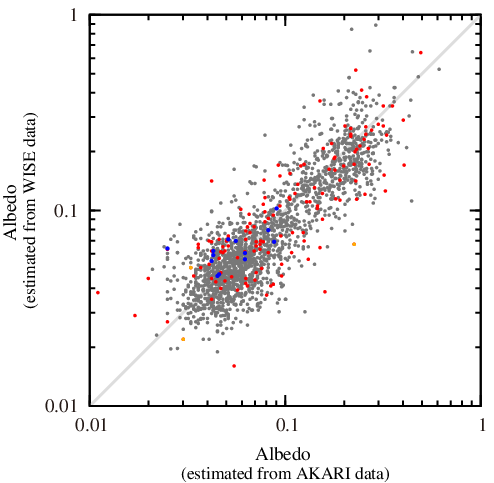}}&%
{\FigureFile(49mm,80mm){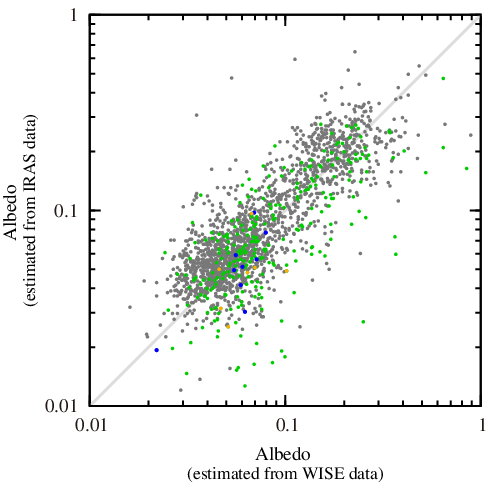}}\\
\\
\hspace{10mm}{\small (d) AKARI vs. IRAS} &%
\hspace{10mm}{\small (e) AKARI vs. WISE} &%
\hspace{10mm}{\small (f) WISE  vs. IRAS} \\
{\FigureFile(49mm,80mm){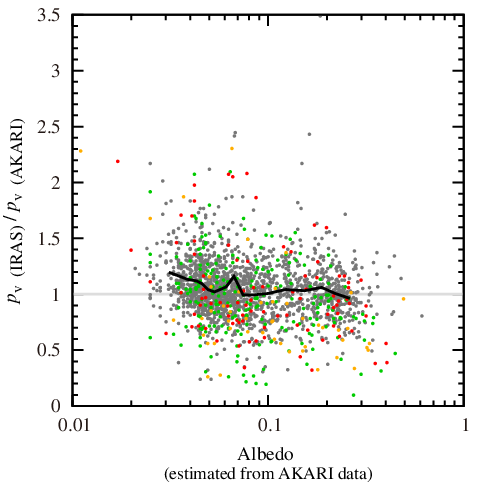}}&%
{\FigureFile(49mm,80mm){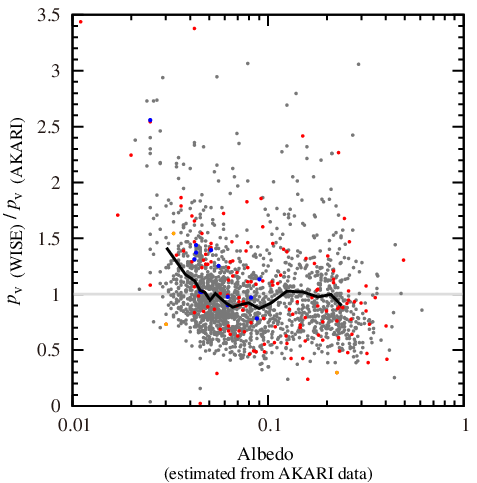}}&%
{\FigureFile(49mm,80mm){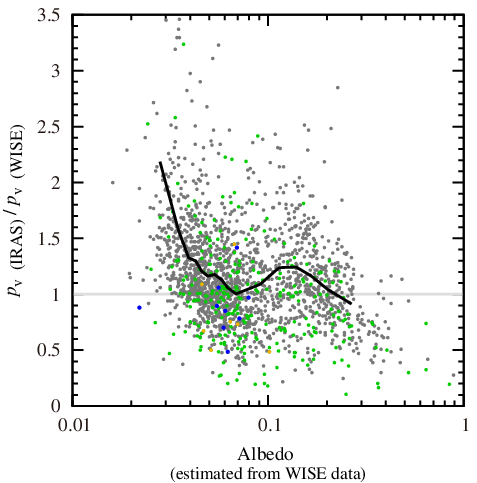}}\\
\end{tabular}
\end{center}
\caption{
Same as figure~\ref{fig: comparing diameter} but in albedo. 
Apart from the small number of detections, 
the mean values of the albedo ratios obtained by the different satellites are 
1.057 (0.296), 
0.986 (0.364), 
and
1.222 (1.216) 
for (d), (e), and (f), respectively, where the numbers in parentheses 
are the standard deviation (1$\sigma$). 
}
\label{fig: comparing albedo}
\end{figure*}

Figure~\ref{fig: comparing diameter} and \ref{fig: comparing albedo} show comparison of the sizes and albedos
of the 1,993 asteroids commonly detected by I--A--W. 
Comparisons between IRAS and AKARI, and between 
IRAS and WISE have been studied~\citep{Usui2011, Mainzer2011c}. 
From these figures, it is evident that the size and albedo measurements by all three satellites are 
in good agreement, although there are some systematic differences. 
In particular, the diameters estimated by {AKARI} are larger than those by {IRAS} but smaller than 
those by {WISE}. 
Therefore, of these three data sets, 
{WISE} yields the largest estimation of size, followed in order by {AKARI} and {IRAS}. 
Conversely, {IRAS} yields the highest albedo, and {WISE} the smallest albedo. 
This inverse relationship between size and albedo is unsurprising given that size 
is inversely proportional to the square root of albedo for a given absolute magnitude 
(e.g., \cite{Fowler1992, Pravec2007}) as: 
\begin{eqnarray}
d &=& \frac{1329}{\sqrt{p_{\rm v}}}~10^{-H/5}\ , 
\label{eq:relation of d and H}
\end{eqnarray}
where $d$, $p_{\rm v}$, and $H$ are the diameter in units of km, the geometric albedo, 
and the absolute magnitude, respectively. 

Figure~\ref{fig:histogram of AKARI, IRAS, WISE commonly} shows 
the distribution of the deviation of the sizes and albedos measured with 
I--A--W, 
from the respective mean values of those three data sets 
for the 1,993 commonly detected asteroids. 
The best-fit Gaussian parameters for these distributions are listed in table \ref{gaussian fitting}. 
As a result, they are in good agreement within $\pm$10\% in diameter and $\pm$22\% in albedo at 
1$\sigma$ deviation level. 

\section{Discussion}
\label{discussion}

\subsection{Survey period of the infrared satellites}

The survey period is one of the important factors for asteroid surveys in the infrared 
(see table~\ref{3 satellites}). At least 
one year is required to survey all of the solar system bodies beyond a semimajor axis of 2~AU 
with the surveyor in a fixed solar elongation of 90$^\circ$, while 
the inertial sky can be covered in half a year. The IRAS mission, which 
lasted ten~months, surveyed approximately 96\% of the sky 
covered with two or more hours-confirming scans~\citep{Neugebauer1984, Beichman1988}. 
The All-Sky Survey conducted for 16 months by AKARI 
fully covered the main belt region, 
using a combination of 170 litres of super-fluid liquid helium and 
two sets of two-stage Stirling cycle mechanical coolers~\citep{Nakagawa2007}. 
Thus, the AKARI asteroid catalog provides a 100\% complete data set of asteroids 
with $H < 9$, as mentioned above. The WISE mission conducted a seven-month-long full cryogenic 
survey and a six-month-long post-cryogenic survey (after depleting its cryogenic tanks). 
In the post-cryogenic phase, only the near-infrared 
channels (3.4 and 4.6~\micron~bands) were used. At these shorter 
wavelengths, asteroid fluxes are a mix of reflected sunlight and 
thermal emissions. Nevertheless, \citet{Mainzer2012}, 
\citet{Masiero2012}, and \citet{Grav2012a} produced reasonable 
estimates of the sizes and albedos of asteroids, assuming a relationship 
between visual albedo and infrared albedo, which are calibrated with data obtained in the full cryogenic phase. 

\subsection{Factors causing discrepancies among IRAS, AKARI, and WISE}

The differences between the mean sizes and albedos obtained by I--A--W are 
generally within $\sim$10\% and $\sim$22\% of each other, respectively (at the 1$\sigma$ standard deviation). 
These values are mostly larger than the uncertainties within each data 
set (typically, 5--13\% for diameter and 10--33\% for albedo). Several possible reasons may explain 
the I--A--W differences. First, the different types of measurements do 
not fully include system uncertainties. Also, several 
factors which may cause discrepancies originate from the physical properties of 
asteroids, such as their shape, thermal inertia, surface roughness, rotation 
rate, and pole orientation. Asteroids are often elongated and 
irregularly shaped, and sometimes form binary systems, which generate 
lightcurve variance as they rotate. The Asteroid 
Lightcurve Database\footnote{http://www.minorplanet.info/lightcurvedatabase.html}
\citep{Warner2009} indicates that, as of September 2013, the mean value of the maximum 
amplitude of the lightcurve for the 5,730 available asteroids is $0.344 \pm 0.296$~mag.
Asteroids with larger amplitude lightcurves are likely to
add to the uncertainty in establishing their size and albedo, especially for estimations 
based on single or a few sightings. 

In addition, uncertainties in the 
treatment of scattered sunlight in the visible wavelengths limit 
the accuracy of radiometric measurements. Usually, simultaneous 
observations in visible and infrared wavelengths are not achieved. 
Instead, the $H$--$G$ system~\citep{Bowell1989} 
is adopted to represent photometric values in visible wavelengths. Uncertainties 
in the absolute magnitudes mainly impact the accuracy of the 
resulting albedo values~\citep{Harris1997}. \citet{Pravec2012} 
found that a discrepancy exists between absolute magnitudes listed in 
the MPC orbit database and those measured by dedicated 
photometric observations over 30 years. They found that the MPC 
values are mostly too small for the 583 observed asteroids; the mean 
offset of $H$ is $-0.4$ to $-0.5$ at $H \sim 14$. 
The slope parameter given in \citet{Pravec2012} varies from $-0.15$ 
to 0.55 (mean, $0.21 \pm 0.09$), while this value is often assumed 
to be $0.15$. These discrepancies can account for the 
uncertainties in the estimated albedo, especially for $H > 10$. 
Once improved measurements of $H$ become available, the values for the sizes and 
albedos can be revised. For example, \citet{Harris1997} devised 
a simple and convenient approximation for recalculating the size and 
albedo from improved $H$ values that does not require detailed thermal model 
calculations. 

Saturation of the observed flux leads to another severe problem for larger 
asteroids. \citet{Lebofsky1989} found that the IRAS observations of 
(1)~Ceres and (2)~Pallas showed unusual behaviors (systematic 
wavelength variations) as compared with the 
results of ground-based observations, perhaps due to 
saturation in 25 and 60~\micron~bands. While these point sources may
be saturated, properly corrected values do not 
affect estimates of the sizes of other objects using the IRAS data~\citep{Tedesco2002}. 
\citet{Cutri2013} reported that point sources detected with 
WISE brighter than 0.88~Jy in 12~\micron~band or 12.0~Jy in 22~\micron~band 
show larger uncertainties owing to the onset of detector saturation. 
The former saturation level corresponds to the thermal emission from 
$\sim$30--70~km sized main belt asteroids. In contrast, no 
sign of saturation is apparent in the AKARI observations~\citep{Ishihara2010}; 
in 18~\micron~band, recorded flux densities for (1)~Ceres were in the range of 500--800~Jy, and 
those for (4)~Vesta were in the range of 470--600~Jy, both of which are below 
the saturation limit (D. Ishihara, 2014, private communication). 

\subsection{The thermal model and the beaming parameter}

The STM (with some modification) or the NEATM can be used to characterize 
the physical properties of asteroids, although care should be exercised 
when applying these simple models to various types of asteroids. The STM 
produces good results if the asteroid has a small thermal inertia, 
rotates slowly, is observed at small solar phase angles, and is not 
heavily cratered or irregularly shaped (i.e., typical larger main belt 
asteroids). However, many asteroids are small irregular 
bodies with predominantly regolith-free rocky surfaces 
and relatively high thermal inertias~\citep{Delbo2007}. Most studies 
support the assumption that asteroid surfaces are generally heavily 
cratered and rough at all scales (e.g., \cite{Ivanov2002}). In 
combination with the lack of an atmosphere and small thermal skin 
depths, surface roughness gives rise to substantial temperature 
contrasts, even at small scales, and produces a {\it beaming effect} 
in which thermal emission is enhanced in the solar direction. 

A beaming parameter was introduced to adjust the surface 
temperature by compensating for the angular distribution of the thermal 
emission (e.g., \cite{Lebofsky1986}). 
The beaming parameter physically correlates with 
the surface roughness and thermal inertia of an asteroid, and in practice, 
it 
can be considered as a normalization or calibration 
factor. For the IRAS catalog, the STM was adopted as the thermal model, 
using a beaming parameter of $\eta = 0.756$, which was derived from 
observations of (1)~Ceres and (2)~Pallas using a ground-based 
telescope~\citep{Lebofsky1986}. It should be noted that the size and albedo 
estimations of 60\% of asteroids (and $> 80$\% of asteroids larger 
than 40~km) detected by IRAS have been revised by a more robust estimation using 
the NEATM, with $\eta$ values ranging from 0.75 to 2.75~\citep{Ryan2010}. The AKARI 
asteroid catalog was processed using the ``modified'' STM, in which 
$\eta$ was determined separately for two observed mid-infrared bands 
($\eta$~=~0.87 for 9~\micron~band and 0.77 for 18~\micron~band), by comparing 
existing data from several different types of measurements for 55 asteroids ranging in diameter from 90 to 960~km 
(see \cite{Usui2011}, table 11). The WISE results used the NEATM with independently 
varying $\eta$ values, which were 
determined by fitting multiple observations using the WISE data alone; 
the results were then examined by comparisons with 49 unique objects with diameters ranging 
from 0.4 to 312~km (data from several sources; 
see \cite{Mainzer2011b}, table 1). The distribution of $\eta$ in the 
WISE thermal model is shown for main belt asteroids in, for example, figure 6 of \citet{Masiero2011} 
(the mean value of their beaming 
parameters is $\eta~=~0.962~\pm~0.153$, for asteroid diameters of $>10$~km). 

Here, we consider the dependency of the size estimation on the value of the beaming parameter, under the conditions of
the thermal model calculation. We assume an asteroid with given 
visible and thermal fluxes at given distances from the Sun and an 
observer. Once an incident solar flux is assumed, an absorbed flux is 
determined (although it depends weakly on albedo). A larger $\eta$ causes 
lower surface temperatures of the thermal flux to balance the absorbed 
flux and the thermal emission. This can be easily found from the 
formulation of the temperature of the subsolar point ($T_{\rm SS}$) 
on the surface of the asteroid (e.g., \cite{Harris1998}), as follows:
\begin{eqnarray}
T_{\rm SS} &=& \left\{\frac{(1-A_{\rm B})S_{\rm s}}{\eta \epsilon \sigma R_{\rm h}^{~2}}\right\}^{1/4}~ ,
\label{eq:max temperature}
\end{eqnarray}
where $A_{\rm B}$ 
is the Bond albedo, $S_{\rm s}$ is the solar flux at 1~AU (i.e., the solar 
constant), $\eta$ is the beaming parameter, $\epsilon$ is the infrared emissivity, 
$\sigma$ is the Stefan--Boltzmann constant, and 
$R_{\rm h}$ is the heliocentric distance in units 
of AU. A lower temperature implies a lower thermal flux 
per unit area. To provide the observed thermal flux, a larger 
asteroid is needed, and a larger size is equivalent to a smaller albedo 
under a given flux in visible wavelengths. Thus, larger $\eta$ values 
reflect larger sizes of asteroids. 

It should also be noted that there is a phase angle (Sun--target--observer 
angle; $\alpha$) dependency of the beaming parameter. In the STM, 
the temperature on the nightside of an asteroid is assumed to be zero, 
which is a reasonable assumption at small phase angles, where 
the dayside flux dominates; i.e., in the case of the main belt 
asteroids. However, \citet{Harris2002} pointed out that care must be exercised 
when applying simple thermal models to the near-Earth asteroids 
because thermal model calculations based on observations made at 
larger solar phase angles are subject to relatively large 
uncertainties; it was because of this that the NEATM was developed~\citep{Harris1998}. 
As compared with main belt asteroids, the near-Earth asteroids tend to have 
irregular shapes, and they are often observed at moderate to large solar 
phase angles ($\alpha > 30^{\circ}$), which is out of the valid range of the STM. 
If the nightside temperature is treated as non-zero, then the relationship between $\eta$ and 
$\alpha$ can vary depending on the temperature distribution. 
Based on the WISE data obtained using the NEATM, the average relationship between 
$\eta$ and $\alpha$ is given as 
$\eta~=~0.00963^{\pm 0.00015} \alpha~+~0.761^{\pm 0.009}$~\citep{Mainzer2011d}, or 
$\eta~=~0.011^{\pm 0.001} \alpha~+~0.79^{\pm 0.01}$~\citep{Masiero2011}, 
although the spread around this value is large for 
$0.3 \leq \eta \leq \pi$. Note that due to the constraints of 
the attitude control of infrared surveyors, the solar elongation 
angle of the WISE observations is fixed at approximately $90^{\circ}$, which means 
that the observing phase angle, heliocentric distance, and geocentric distance 
are strongly correlated. 

The thermophysical model (TPM, \cite{Lagerros1996, Lagerros1997, Lagerros1998}), 
which is a sophisticated approach for asteroid modeling, assuming a spherical 
body, is developed to derive the size, albedo, thermal inertia, and 
sense of rotation without assuming a value of the beaming parameter. 
\citet{Mueller2014} discussed the validity of the TPM for a selected 
target by using only this simple spherical shape model. 

\subsection{Comparisons with the other measurements}

Importantly, none of the I--A--W catalogs consider 
the irregular shapes of asteroids. While a non-rotating spherical body is 
assumed for asteroids in both the STM and NEATM, the actual shapes of 
asteroids are generally elongated, especially in the cases of smaller asteroids. 
Figure~\ref{fig:relative difference} shows the relative differences 
between the diameters measured by I--A--W and the effective (volume-equivalent) 
diameters ($D_{\rm ref}$) derived from the shape models determined by 
several methods: direct imaging with the Hubble Space Telescope~\citep{Tanga2003}, 
with the adaptive optics system on the W.M. Keck II telescope 
\citep{Hanus2013, Marchis2006, Drummond2009, Conrad2007}, or 
by spacecraft observations\footnote{Although more than ten asteroids have been explored by spacecraft %
flyby/rendezvous/landing/sample return, only (243) Ida ($d \sim 30$~km) %
has also been observed by all three infrared surveyors.}%
\citep{Thomas1996}, stellar occultation combined with lightcurve 
inversion techniques~\citep{Durech2011}, speckle 
interferometry~\citep{Cellino2003, Drummond1985}, and radar 
observations~\citep{Ostro2000}. In total, 88 main belt asteroids 
ranging in size from 30 to 540~km are included. The relative difference is 
defined as $\left(D_{i} - D_{\rm ref}\right) / D_{\rm ref}$, where 
$i$ refers to IRAS, AKARI, or WISE. The mean values of the relative 
differences are 2.8\%, 1.7\%, and 7.5\% for IRAS, AKARI, and WISE, 
respectively, and the standard deviations for each are 12--13~\%. 
We found that the size derived by AKARI is closer to that derived 
by IRAS or WISE. This is not a surprising 
result, as the beaming parameter adopted in the thermal model 
calculation in the AKARI catalog is calibrated with 
well-studied main belt asteroids larger than 90~km, whose size, 
shape, rotational properties, and albedo are known from different 
measurements, as mentioned above. In this respect, the diameters 
obtained by radiometric measurements based on I--A--W are reliable 
in a statistical sense, which are smoothed out and averaged over 
a limited number of observations, even though the sizes obtained by radiometric and other measurements can be 
discrepant by up to 30\%. 

\section{Summary}

A total of 138,285 asteroids were detected with three infrared all-sky surveyors: 
IRAS, AKARI, and WISE, which enabled their sizes and albedos 
to be determined by the radiometric method. 
IRAS made a pioneering asteroid survey including 2,470 objects. 
AKARI resulted in a 100\% complete survey for larger asteroids 
($H < 9$, corresponding 40~km or larger) in its 16-month mission. 
WISE has significantly improved the number of smaller sized asteroids detected 
($H < 15$, corresponding 1.5~km or larger) as a result of its higher sensitivity compared with IRAS and AKARI. 
1,993 asteroids were commonly detected by all three satellites, 
and the size and albedo measurements of these asteroids by each satellite are in good agreement 
(within $\pm$10\% for diameter and $\pm$22\% for albedo). 
The data sets from these three satellites complement one another and will provide 
an important database for statistical analysis of asteroid populations and characterizing 
the physical properties of each asteroid. 

\bigskip

This study is based on observations with AKARI, a JAXA project with 
the participation of ESA. 
This work also makes use of data products from the Wide-field Infrared Survey Explorer, 
which is a joint project of the University of California, Los Angeles, and 
the Jet Propulsion Laboratory/California Institute of Technology, funded 
by the National Aeronautics and Space Administration. 
FU would like to thank Takao Nakagawa (ISAS/JAXA), Daisuke Ishihara (Nagoya University), and 
Takashi Onaka (the University of Tokyo), for their valuable comments. 
MI is supported by the National Research Foundation of Korea (NRF) grant
funded by the Korea Government (MEST) (No. 2012R1A4A1028713).
SH is supported by the Space Plasma Laboratory, ISAS/JAXA. 
TO is supported in part by JSPS KAKENHI Grant Number 25400220.
We thank the anonymous referee for careful reading and providing constructive suggestions. 

\clearpage

\clearpage

\clearpage

\clearpage

\clearpage

\clearpage

\begin{figure*}
\begin{center}
\hspace*{-1cm}
\begin{tabular}{c}
{\FigureFile(100mm,80mm){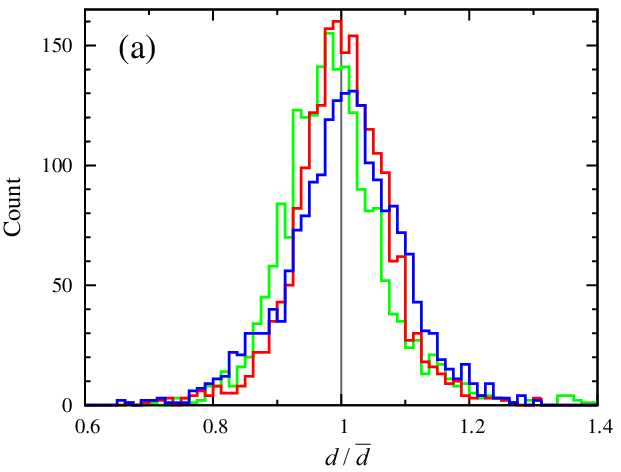}}\\
{\FigureFile(100mm,80mm){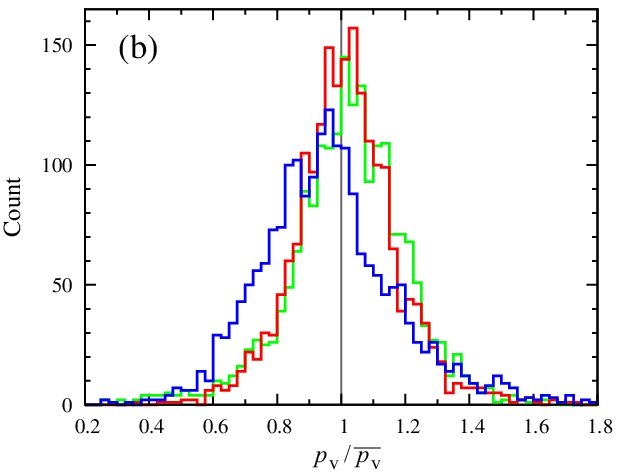}}\\
\end{tabular}
\end{center}
\caption{
The distribution of the deviation of 
diameter ($d$) and albedo ($p_{\rm v}$) measured 
by IRAS (green), AKARI (red), or WISE (blue), 
from the mean values of the three satellites ($\overline{d}$, $\overline{p_{\rm v}}$)
for the 1,993 commonly detected asteroids. 
The means and standard deviations of the
best-fitting Gaussian curves are summarized in table \ref{gaussian fitting}.
}
\label{fig:histogram of AKARI, IRAS, WISE commonly}
\end{figure*}

\clearpage

\begin{figure*}
\begin{center}
\hspace*{-1cm}
{\FigureFile(150mm,80mm){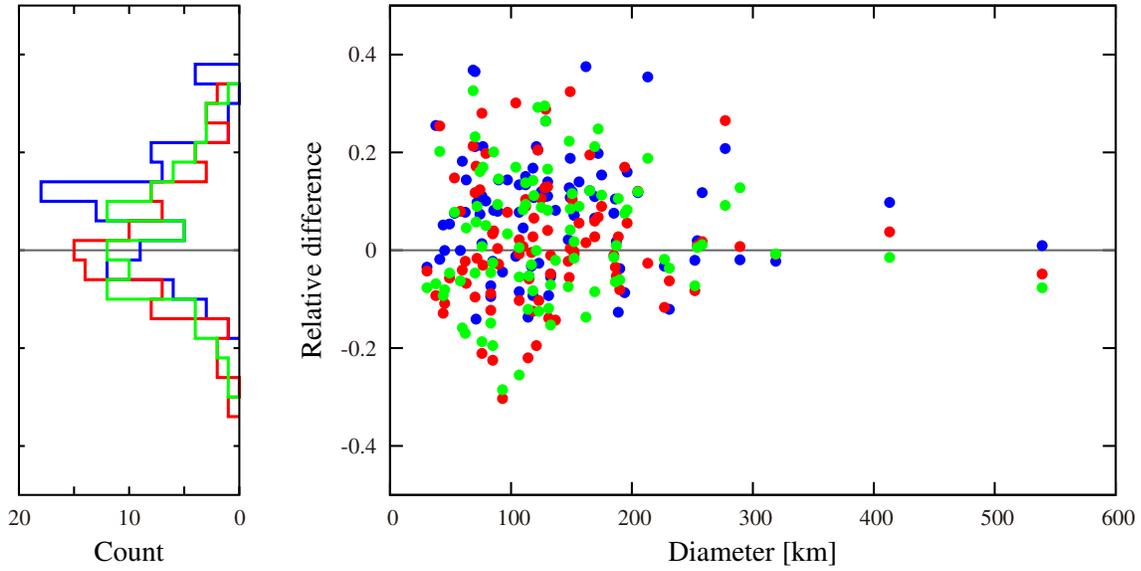}}
\end{center}
\caption{
Relative difference between effective diameters ($D_{\rm ref}$) determined by direct imaging with HST or Keck AO, 
occultations, speckle interferometry, or radar (see text) and diameters derived by I--A--W. Here relative difference 
is defined as $\left(D_{i} - D_{\rm ref}\right) / D_{\rm ref}$, where $i$ refers to IRAS, AKARI, or WISE. 
Color dots mean the diameter measured by IRAS (green), AKARI (red), or WISE (blue).
}
\label{fig:relative difference}
\end{figure*}

\clearpage

\begin{table}
\caption{Details of the infrared all-sky survey satellites. \label{3 satellites}}
\begin{tabular}{lccc}
\hline
& IRAS$^{a}$ & AKARI$^{b}$ & WISE$^{c}$\\
\hline
Size [m] & $3.6 \times 3.2 \times 2.1$  & $5.5 \times 1.9 \times 3.7$ & $2.9 \times 2.0 \times 1.7$ \\
Weight [kg] & 1083 & 952 & 750\\
Telescope & 57cm, $f$/9.56  & 68.5cm, $f$/6 & 40cm, $f$/3.375 \\
          & Ritchey-Chr\'{e}tien & Ritchey-Chr\'{e}tien & Mirror complex$^{d}$\\
Cryogen   & LHe (480 $\ell$) & LHe (179 $\ell$) with & Solid hydrogen (15.7 kg) \\
          &             & 2-stage Stirling coolers & \\
Altitude of orbit & 900 km & 750 km & 525 km\\
Launch [UT]& 1983/01/26~02:17:00 & 2006/02/21~21:28:00 & 2009/12/14~14:09:00\\
End of operation& 1983/11/23 & 2011/11/24~08:23:00 & 2011/02/17~20:00:00$^{e}$\\
\multicolumn{1}{c}{[UT]}\\
Period of survey$^{f}$ & 1983/02/09--1983/11/22 & 2006/04/26--2007/08/26 & 2010/01/07--2010/08/05\\
                                  & 287 days & 488 days & 211 days\\
Wavelengths & 12,25,50,100 & 9,18,65,90,140,160$^{g}$ & 3.4,4.6,12,22\\
\multicolumn{1}{c}{[\micron]}\\
5$\sigma$ sensitivity & 350,330,430,1500$^{a}$ & 50,90$^{h}$,3200,550,3800,7500$^{i}$ & 0.08,0.1,0.85,5.5$^{j}$ \\
\multicolumn{1}{c}{[mJy]} \\
FOV$^{k}$ [$^{\prime}$~] & 4.5$\sim$5.0 & $\sim$10 & 47\\
Sky coverage & $>$ 96\% & $>$ 96\% & $>$ 99\%\\
\hline
\\
\multicolumn{4}{l}{\hbox{\parbox{148mm}{\footnotesize
      \par\noindent
      ({a}){The Infrared Astronomical Satellite, ~\citet{Neugebauer1984, Beichman1988}}
({b}){\citet{Murakami2007}; 
pre-launch designation is ASTRO-F. 
Note that AKARI means ``light'' in Japanese and is not assigned a special acronym.}
({c}){The Wide-field Infrared Survey Explorer, ~\citet{Wright2010, Cutri2013}}
({d}){The optical sub-assembly of WISE consisted of an afocal telescope with six mirrors, 
a scan mirror, 
and imaging optics with six mirrors, which are required to cover wide field-of-view and have extremely 
low distortion for performing internal scanning without image blurring. 
It is a kind of ``a Cassegrain-like objective''~\citep{Schwalm2005}.
}
({e}){There is a report that WISE would be reactivated in September 2013.}
({f}){Cryogenic cooled phase only.}
({g}){While the all-sky survey with AKARI was measured in six bands, 
the asteroid catalog was constructed using two mid-infrared bands (9 and 18 \micron).}
({h}){\citet{Ishihara2010}}
({i}){\citet{Yamamura2010}}
({j}){\citet{Mainzer2011a}}
({k}){In scan direction.}
      }\hss}}
\end{tabular}
\end{table}

\clearpage

\begin{table}
\caption{Best-fit parameters of the Gaussian curve fitting for the size ($d$) and albedo ($p_{\rm v}$) distributions 
in figure~\ref{fig:histogram of AKARI, IRAS, WISE commonly}. \label{gaussian fitting}}
\begin{center}
\begin{tabular}{lllll}
\hline
 & \multicolumn{1}{c}{$\overline{d}$} & \multicolumn{1}{c}{$\sigma(\overline{d})$} & \multicolumn{1}{c}{$\overline{p_{\rm v}}$} & \multicolumn{1}{c}{$\sigma(\overline{p_{\rm v}})$}\\
\hline
IRAS  & 0.992 & 0.094 & 1.022 & 0.193\\
AKARI & 1.001 & 0.076 & 1.016 & 0.164\\
WISE  & 1.006 & 0.093 & 0.962 & 0.222\\
\hline
\end{tabular}
\end{center}
\end{table}

\end{document}